\documentclass[doublecol]{epl2} 
\usepackage{epstopdf}
\usepackage{amsmath}
\usepackage{color}

\title{Fidelity, entropy, and Poincar\'e sections as tools to study the polyad breaking phenomenon}
\shorttitle{Tools to study the polyad breaking phenomenon} 

\author{M. Berm\'udez-Monta\~na\inst{1} \and R.Lemus\inst{1} \and O.Casta\~nos\inst{1}}
\shortauthor{R.Lemus\etal}

\institute{                    
  \inst{1} Instituto de Ciencias Nucleares, 
Universidad Nacional Aut\'onoma de M\'exico,\\   Apartado Postal 70-543, 
04510 M\'exico, DF, M\'exico. \\
}

\abstract{In search of a region where a local mode model stop being adequate to estimate the local force constants, the correlation diagram of the vibrational energy spectra associated with the stretching modes of triatomic molecules such as CO$_2$ and H$_2$O is analyzed  by means of two interacting Morse oscillators. By considering a linear dependence of the structure and force constants, it is shown that the fidelity, entropy and Poincar\'e sections detect the polyad breaking process manifested in the 
transition from   local to   normal  mode behaviors. Additionally Poincar\'e sections show a transition to chaos  where the local polyad cannot be defined.}

\usepackage{graphicx}
\graphicspath{%
    {converted_graphics/}
    {/}
}
\begin{document}
\maketitle
\section{Introduction}
In the description of molecular vibrational excitations the normal modes (NM) picture has  played a fundamental role. At first only diagonal interactions are considered but latter on resonances may be  included through the use of the concept of  polyad, a pseudo quantum number that encompass all the eigenstates connected with the relevant interactions \cite{kellmanpolyad}. This  analysis  is  conveniently treated in an algebraic scheme introducing bosonic operators associated with the NM \cite{praga,kellman}.  The patterns may be modified either by  low  potential barriers \cite{praga,jensen}, or by the appearance of nearly degenerate states, a signature of molecules with local character \cite{ lawton,jensen2,child,halonen2p}.

In a local model (LM) the Hamiltonian is expressed in terms of a set of interacting  local oscillators, a model that provides a reasonable description  when large mass differences are present with the remarkable property that a polyad can be defined. Even in molecules with LM behavior a NM  behavior may be manifested. The study of local-to-normal mode behaviors is of interest because of their connection to intramolecular vibrational energy transfer and their possible role in facilitating or inhibiting reactivity. This fact has stimulated quantum mechanical studies \cite{lawton,jensen2,child,halonen2p,sako1,sako2}, but  incorporating also modern
  methods of nonlinear classical mechanics
 \cite{sibert,davis,heller,kellmanacc,kellman88,coy,kellman89,Sako,xiao}.

\textcolor{red}{If we consider vibrational  levels of a molecule in the medium energy range   with an extreme LM behavior, these are characterized by a polyad multiplet value $P_L$ (total number of local quanta associated with each oscillator). By allowing an interaction strength between the two oscillators to be increased, the levels split and start approaching together leading to a normal behavior.} When the splitting   become  so important that levels of different polyads approach and even cross, a local mode model stop being appropriate \cite{jensen2}. In this case a normal mode  scheme NM with polyad  $P_N=\nu_1+\nu_3$~
 should be more convenient from the outset {\it e.g.} CO$_2$. \textcolor{red}{These extreme behaviors manifest through the connection between the  polyads defined in the normal  and  the local  schemes, which takes the general form $P_N=\zeta_0+\beta_0 P_L+\alpha \hat V$, where $\hat V$ is a  contribution not preserving $P_L$. The parameters $\alpha,\zeta_0$ and $\beta_0$ depends upon the force and structure constants in such a way that $\alpha,\zeta_0 \to 0$  and $\beta_0 \to 1$ for molecules with local mode behavior. In this contribution the study of the local-to-normal mode transition (LNT) is presented in the framework of this relation between polyads.}

There are several concepts that may be used to identify sudden changes in a quantum state. The probability density for instance reflect the degree of locality, but a more sensitive functions are fidelity and entropy.
The fidelity and Shannon entropies are concepts introduced in the classical information theory. The first measures the accuracy of a transmission message while the second one is related with the {\it  coding theorems}, i.e., how much can be compressed a message without losing information~\cite{casati}. These concepts were extended to quantum information theory. The fidelity is used to compare quantitatively two probability distribution functions which for pure states is related to the overlap of two quantum states. The von Neumann entropy plays an   analogous role to the Shannon entropy for quantum channels. Additionally for bipartite systems it measures the degree of entanglement of the components of the system.  The fidelity concept has also been used to determine the quantum phase transitions of the ground state of a quantum system when a parameter of the Hamiltonian is changed continuously~\cite{jpa2012, romera}. As in the quantum phase transition there is a sudden change in the properties of the ground state; it has been found that the von Neumann entropy takes extremal values \cite{jpa2012}. In addition to the  fidelity and Shannon entropy, modern methods of non linear classical mechanics may also be helpful in the identification of the phase transition. In particular the Poincar\'e sections will be used in order to identify chaos,  corresponding   to  regions where polyads are not preserved and representing possibly  unstable situations.

 In this work we address the problem of studying  the transition from a molecule that can be described in a local scheme to a molecule whose local mode description is  unfeasible unless the polyad $P_L$ is broken. A fundamental issue consists in identifying the relevant physical parameterization.  As a reference  we  consider the limit systems H$_2$O and  CO$_2$.  In our analysis the concepts of probability density, fidelity, entropy as well as Poincare sections represent our tools to identify the transition,  whose results are presented for a specific set of eigenstates.

The present work is organized as follows.  First the basic features of LM and NM behaviors are revisited, providing the relevant parameters that allow the identification of the LNT.
Thereafter we study the LNT using interacting Morse oscillators as a model for the molecular vibrational excitations. This transition is analyzed with assistance  of quantum mechanical concepts as well as with aid of classical  mechanics through the construction of Poincar\'e sections. Finally a  summary and concluding
remarks are presented.

\section{Relevant parameters involved in the  local-to-normal mode transition}

The vibrational Hamiltonian for a set of two equivalent oscillators  presenting  LM behavior can be written in  the form
$
\hat H_{L}=\hat H^{(loc)}_0+\hat V^{(loc)}_{int},
$
where $\hat H^{(loc)}_0$ corresponds to two non interacting local oscillators, while $V^{(loc)}_{int}$ involves interactions  presumably  playing the role of a perturbation, yet fundamental in the physical description. A sensible way to construct $\hat V^{(loc)}_{int}$  consists in identifying resonances  preserving the polyad $P_L$, which consists in the total number of local quanta associated with each oscillator. This is justified by the fact that  at least in the low energy region of the spectrum the general feature of the spectrum consists of a  well separated set of closed   levels characterized by $P_L$.
In contrast, when the masses are similar \textcolor{red}{and the geometry linear}
 a more convenient starting point for the Hamiltonian may be  a NM scheme defined by
$\hat H_{N}=\hat H_0^{(nor)}+\hat V_{int}^{(nor)}$,  
where now $\hat H^{(nor)}_0$ corresponds to the set of non interacting harmonic oscillators and  $\hat V_{int}^{(nor)}$ involves  diagonal    as well as resonant interactions preserving  $P_N$. 

The analysis of the vibrational spectroscopy starts by identifying the fundamentals, from which the polyad $P_N$ is determined. The polyad is expected to be a good quantum number 
as we remain in the low region of the spectrum. As the energy increases anharmonic effects become manifest  breaking the polyad. \textcolor{red}{Considering that a local description is derived from local coordinates and normal description from normal coordinates}, in  general  $P_L \neq P_N$. However for molecules with  local character the transformation reduces to a canonical transformation and $P_L \approx P_N$. In practice this permits to write down the polyad preserving Hamiltonian in a local representation in a straightforward way. From the spectroscopic point of view such situations are present because the energy splitting due to the  interaction between the local  oscillators is  considerable lesser than the distance between groups of levels associated with different polyads $P_L$. \textcolor{red}{As the interaction increases},  a mixing of states with different polyads $P_L$ appears, 
ending with  only  $P_N$ as a good quantum number. The lost of the quantum number $P_L$ suggest a transition region based an a polyad breaking process  connected with the feasibility of a local mode treatment. 

According to our knowledge, LNT has not been  studied  from the perspective \textcolor{red}{ of local polyad breaking}. The traditional analysis is only concerned with molecules presenting a LM behavior ($P_L=P_N$) and the degree of locality $\xi=\frac{2}{\pi} \arctan{{\lambda \over \omega x}}$ refers to the splitting of the levels due to the interaction of the oscillators (parameter $\lambda$) relative to the intensity of the anharmonicity (parameter $\omega x$) \cite{jensen2,child,halonen2p}. Hence the analysis is focused upon the splitting of a multiplet characterized by $P_L$. Because of the identity $P_L=P_N$, the same Hamiltonian can be described to a NM scheme, which makes the descriptions to be equivalent through the $x$-$K$ relations and classical trajectories in phase space  \cite{kellman89,mills}.

We present now a  novel  analysis in which we consider the transition between two molecular systems strictly  characterized by LM and \textcolor{red}{ NM} behaviors, water and carbon dioxide, for instance. Because $P_L$ \textcolor{red}{ is only  preserved  }   in H$_2$O, the transition to CO$_2$ involves a polyad breaking process with  conspicuous changes in the molecular properties. A fundamental ingredient for this analysis is the parameterization used to connect the systems, \textcolor{red}{ motivated from previous works on the description carbon CO$_2$ using an algebraic local model}  \cite{co2}. The parameterization comes from the analysis of two interacting harmonic oscillators up to quadratic terms, whose \textcolor{red}{ local} description in second quantization takes the form
\begin{align}
\label{hama12}
 \hat H={\hbar \omega \over 2} \hat H^{H.O}+\lambda (\hat a_1^\dagger  \hat a_2+ \hat a_1  \hat a_2^\dagger)  +\lambda' (\hat a_1^\dagger \hat  a_2^\dagger+ \hat a_1 \hat a_2), 
\end{align}
where \small$ \hat H^{H.O}=\sum_{i=1}^2(\hat a_i^\dagger \hat a_i+ \hat a_i  \hat a_i^\dagger)$\normalsize, with \small$\omega=\sqrt{f_{rr} g^{{\tiny \mbox{o}}}_{rr}}, \lambda={\hbar\omega\over 2}\left( x_f+x_g   \right)$, \normalsize and \small $\lambda'= {\hbar \omega \over 2} \left( x_f-x_g \right)$\normalsize, where  $x_f=f_{rr'}/f_{rr}$
and $x_g=g^{{\tiny \mbox{o}}}_{rr'}/g^{{\tiny \mbox{o}}}_{rr}$.
 The Hamiltonian (\ref{hama12}) does not preserve  $P_L$, \textcolor{red}{ unless  the last term in (\ref{hama12})  is negligible. In such case  the system is identified with  a LM behavior}. The Hamiltonian
with $\lambda'=0$  becomes the basic model to describe molecules like water although anharmonicities must be incorporated in order to reproduce the experimental energy splitting \cite{lawton}. Since the Hamiltonian  (\ref{hama12}) is integrable it  may be put in the form $\hat H = \hbar \omega_ {{g}}  \hat \nu_1 + \hbar \omega_ {{u}} \hat \nu_3$
with frequencies \small $\omega_ {{g}}= \omega \sqrt{(1+x_f) (1+x_g)}, \, \omega_ {{u}}= \omega \sqrt{(1-x_f) (1-x_g)}$\normalsize. The Hamiltonian in \textcolor{red}{normal coordinates} is diagonal in the normal basis $|\nu_1 \nu_3 \rangle $.  The Hamiltonian preserves the polyad $P_N$ in both representations, but $P_L$ is preserved only by (\ref{hama12}) when $\lambda'=0$. A fit of experimental energy levels may be achieved with any of the two Hamiltonians providing the same results, and from them we may extract the force constants. The question which arises is concerned with appropriate values of the structure and force constants  that allows the Hamiltonian (\ref{hama12}) to be used with $\lambda'=0$ for computing  the force constants. This is equivalent to establish the limit values of ($x_f$,$x_g$) for a molecule to be considered local. It has been proved that this  condition is \cite{co2}
\begin{equation}
\label{condi12}
\gamma \equiv {1 \over 8}(x_f-x_g)^2 \ll1.
\end{equation}
  Since $\gamma$ is associated with the splitting of the fundamentals relative to  themselves, we find convenient to introduce the parameter
\begin{equation}
\zeta=\frac{2}{\pi}\arctan \left(\frac{\Delta E }{\bar E}\right),
\end{equation}
 where  $\Delta E$ stands for the difference between the fundamentals $\Delta E=\nu_1-\nu_3$, while $\bar E$ corresponds to their average $\bar E=(\nu_1+\nu_3)/2$. In Fig. 1 we display the location of several molecules in a plot  $\zeta$ vs $\gamma$. Hence  a normal mode region is identified with the upper right part, while the local mode zone with  the lower left part of the plot, with an intermediate region  closely related with the  breaking of the polyad $P_L$. This diagram suggests a study of the parametric change  from a local molecule like H$_2$O to a normal molecule like CO$_2$ as indicated with an arrow.  Along this line we expect to identify a transition region, albeit
considering  two interacting Morse oscillators.

\begin{figure}[h]
\begin{center}
\includegraphics[width = 0.66\columnwidth]{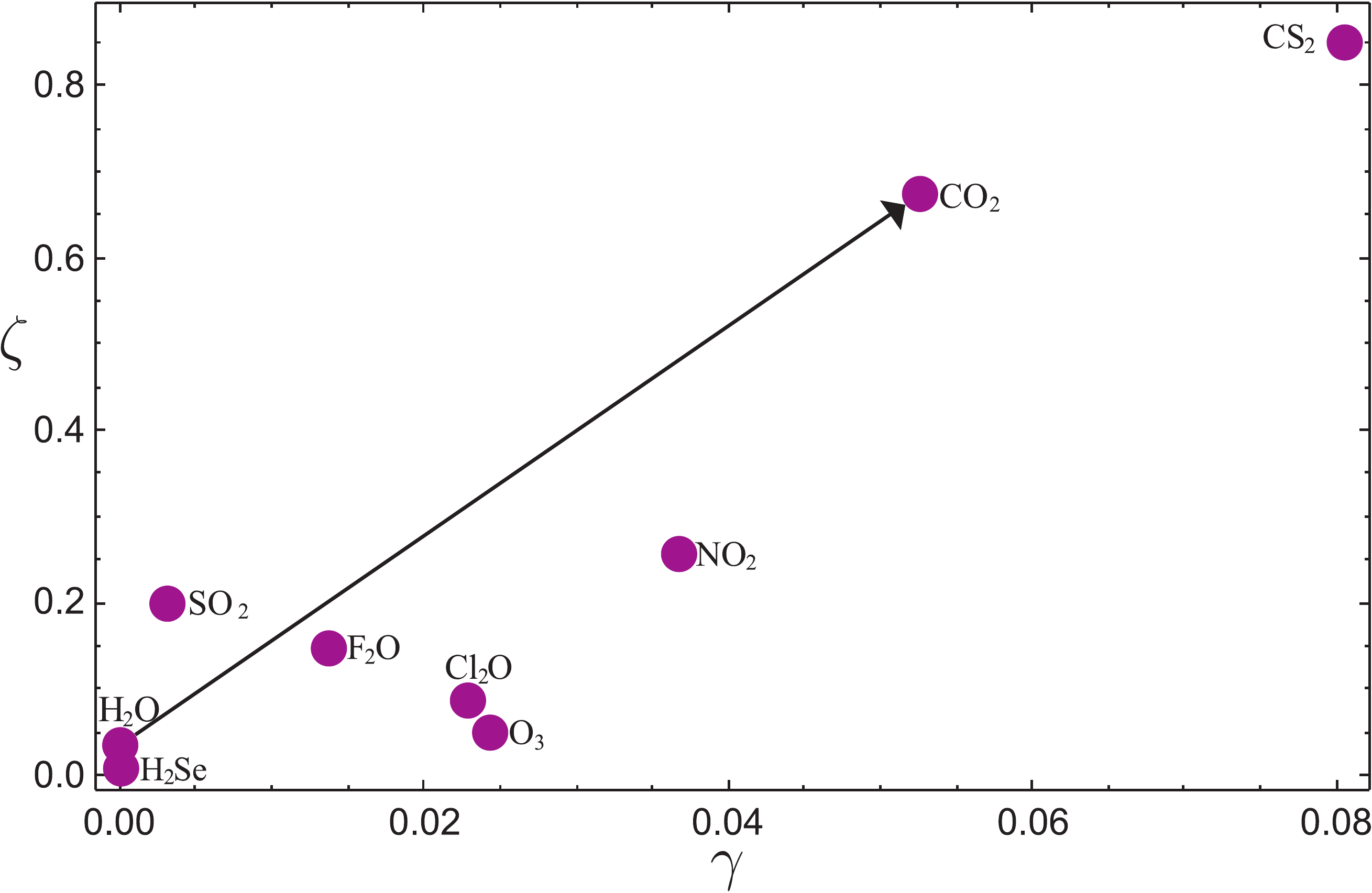}
\end{center}
\caption{Location of several molecules  in the diagram  $\zeta$ {\it vs} $\gamma$. The arrow indicates the parameterization considered to study the LNT.}
\end{figure}

\section{Local-to-Normal mode transition}

In this section we shall analyze the LNT through the study of the stretching modes of a triatomic molecule modelled with a  Hamiltonian of two interacting Morse oscillators. The Morse potential can be expressed as \small$V_M(q)=D y^2$ \normalsize with \small$y=1-e^{-\beta q}$\normalsize. The number of quanta $v_i$ for each oscillator takes the values $v_i=0,1, \dots, j-1$ with $\kappa=2 j+1$ related with the depth of the potential $D$. We introduce a linear parameterization in the $\{ x_g \equiv t,x_f(t) \}$ space from the water parameters \small$(x_g^{(L)}=-0.015,  x_f^{(L)}=-0.012)$ \normalsize  to the ones associated with the carbon dioxide \small$(x_g^{(N)}=-0.571,x_f^{(N)}=0.047)$\normalsize. This parameterization induces the  linear $t$-dependence for the frequency \small$\omega(t)=m_\omega t+w^{(N)}$\normalsize and the Morse parameter \small$\kappa(t)=m_\kappa t+\kappa^{(N)}$\normalsize, where \small$m_\omega=(\omega^{(N)}-\omega^{(L)})/( x_g^{(L)}-x^{(N)}_g), m_\kappa=(\kappa^{(N)}-\kappa^{(L)})/( x_g^{(L)}-x^{(N)}_g)$\normalsize, with \small
$k^{(L)}=48$, $k^{(N)}=160$\normalsize, and \small$\omega^{(L)}=1853$ cm$^{-1}$, $\omega^{(N)}= $ 959~ cm$^{-1}$\normalsize, which were chosen in order to reproduce the fundamentals. The $\beta(t)$ is given by \small$\beta(t)=\sqrt{[2w(t)]/[\hbar \kappa(t)g_{rr}^{\circ}(t)]}$\normalsize. Then the Hamiltonian for the two interacting Morse oscillators takes the form
\begin{align}
\label{2Mot}
\hat H (t)  =&\hbar \omega(t) \bigg\{ \sum_{i=1}^2  \big[ (\hat v_i+1/2)-{1 \over \kappa(t)}(\hat v_i+1/2)^2 \big] \nonumber \\&+ {2 ~t \over \kappa(t)} \hat {\bar p}_1 \hat {\bar p}_2+{\kappa(t)~ x_f(t) \over 2} \hat y_1 \hat y_2 \bigg\},
\end{align}
where here the momenta $\hat {\bar p}_i$ are dimensionless.  We should stress that the space $|jv_1 v_2 \rangle$ is divided into two subspaces, the one belonging complete polyads and the rest belonging to the continuum \cite{osiris}.

The parameterization  $\kappa(t)$    implies  different dimensions for the Hamiltonian matrix representation. Since we are interested in the low lying region of the spectrum  we have kept the dimension constant (consistent with  $\kappa=10$), albeit
changing $\kappa$ in accordance  with $\kappa(t)$  in the calculation of the matrix elements. In this way we simplify the numerical description without losing physical content.  In Fig. 2 we display the energy correlation diagram for the first $42$ symmetric eigenstates provided by the Hamiltonian (\ref{2Mot}).  The left hand side corresponds to the local limit where a clear polyad preserving pattern is evident up to polyad $6$.   In this limit the polyad in terms of local and normal number operators coincide $P_N=P_L$.   As $|t|$ increases apparent level crossings appear  suggesting  the location of the LNT. We will show however that the transition is not determined by these crossings, but by properties carried by the eigenstates as the polyad $P_L$ is broken.  There are several sensitive properties   that provide  a precise information for the transition region, on which we base our strategy:

 a) Components. The analysis of the dominant components of the eigenkets in both local and normal basis should reflect the transition. The normal basis, however,  deserves some discussion since strictly speaking  a normal basis does not exist in a set of Morse oscillators. In order to extract from  the eigenstates the components of the normal basis we construct the normal states diagonalizing the number operators $\{\hat \nu_1,\hat \nu_3 \}$ in the harmonic local basis $|n_1 n_2 \rangle$. The resulting transformation matrix is inverted to substitute the local basis in the Morse eigenstates with the following identification $|n_1 n_2 \rangle \to |jv_1 v_2 \rangle$. This approach is feasible as long as the maximum component  of the  eigenstates is located in the subspace of complete polyads.

b) Fidelity. Another property to extract information about the transition is  
 through the fidelity $F_\alpha(t)$ associated with a given   eigenstate $|\Psi_\alpha \rangle$, and defined as the overlap between consecutive eigenstates parametrically separated by $\delta t$: \small
$
F_\alpha(t)=|\langle \Psi_\alpha;t| \Psi_\alpha;t+\delta t\rangle|^2.
$\normalsize

c) Entropy. The transition may also be manifested  through the  entanglement between the two oscillators,  a quantitative property measured calculating the  entropy defined as \cite{casati}:
$
S_\alpha=-\sum_i \lambda_i \ln \lambda_i,
$
where $\lambda_i$ is the $i$-th eigenvalue of the matrix \small$|| \langle j v_1'|Tr_2 \rho_\alpha |jv_1 \rangle||$ \normalsize with density operator \small$\rho_\alpha=| \Psi_\alpha \rangle \langle \Psi_\alpha |$\normalsize, while \small$Tr_2 \rho_\alpha=\sum_{v_2} \langle j v_2|\Psi_\alpha \rangle \- \langle \Psi_\alpha|j v_2 \rangle$\normalsize. In the local limit the entropy vanishes, and it increases as the coupling appears. 

d) Probability density. We may also see the transition by plotting the probability density $\rho_\alpha$ associated with the eigenstate $| \Psi_\alpha \rangle$ in the coordinate representation: \small $\rho_\alpha(q_1,q_2)=|\langle q_1 q_2|\Psi_\alpha \rangle|^2$\normalsize. This property has proved to be useful in reflecting the local-normal character \cite{child}.

e) Poincar\'e sections. Since during the LNT the polyads $P_L$ and $P_N$ are not preserved, chaos is expected to appear  \cite{kellman97,ezra97}.  Consequently  chaotic phase spaces in the polyad breaking regions should be manifested. In order to identify the transition, Poincar\'e sections will be relevant. 

\section{Analysis of the results}
\begin{figure}[h]
\begin{center}
\includegraphics[width = 0.8\columnwidth]{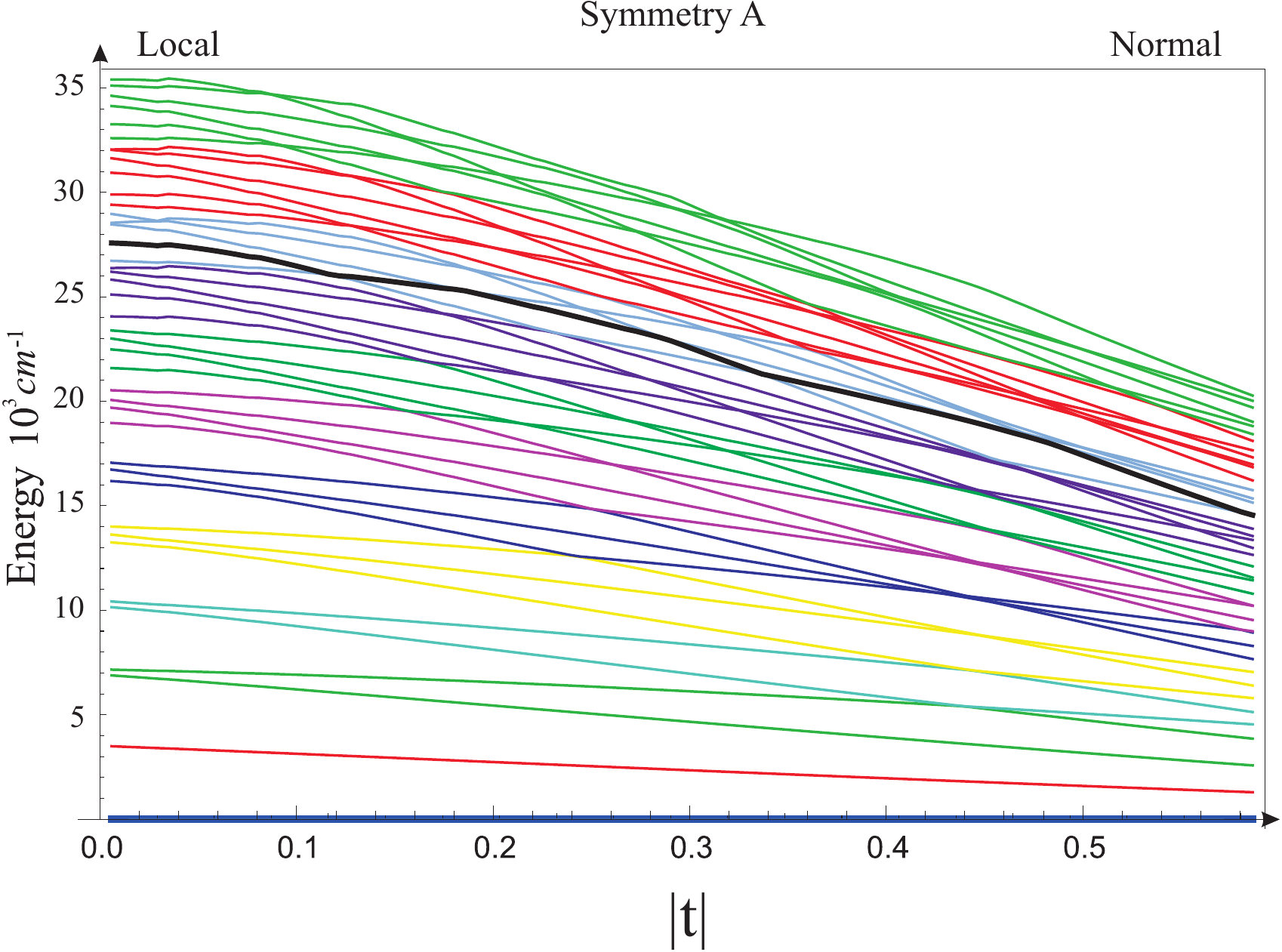}
\end{center}
\caption{Energy correlation between the local and normal limits for the symmetric states for two interacting Morse oscillators.}
\end{figure}
We have chosen the states $\{26,27,28   \}$ as a  representative set displaying the main features of the energy spectrum  by the set of interacting Morse oscillators. These states can be located in Fig. 2, where the state $27$ stands out in black. Although the whole range of the transition from water to carbon dioxide is displayed, we shall constraint our analysis in  the upper left part of the spectrum in the interval $|t| \in [0.0147,0.2]$, since it is in this region 
where the LNT is manifested in  different forms. In Fig. 3a a zoom of the three levels is shown. Although at first sight crossing of levels appears, a more detailed analysis shows that they are   avoided crossings \cite{hari,gbo}.

\begin{figure}
\begin{center}
\includegraphics[width = 0.72\columnwidth]{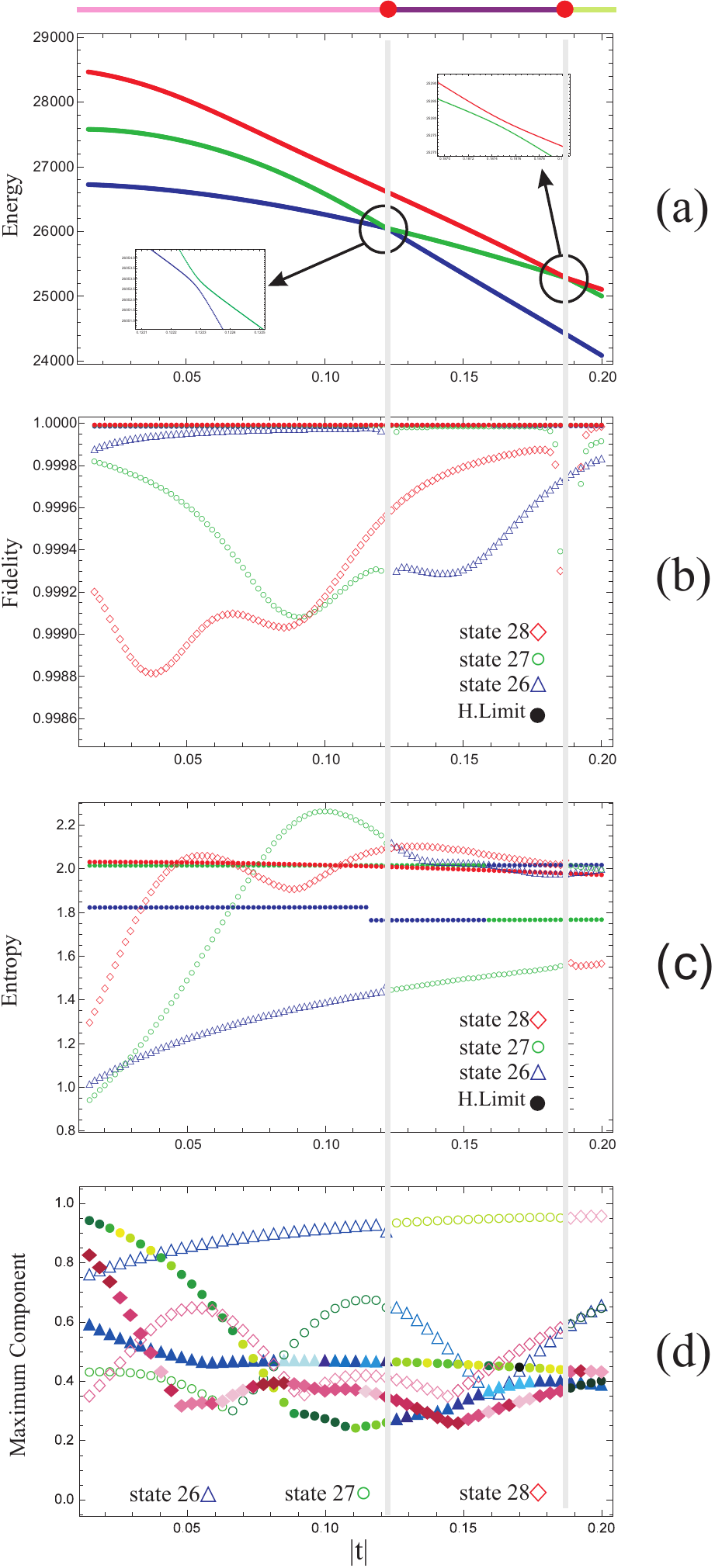}
\end{center}
\caption{Properties of the set of states $\{ 26,27,28 \}$ in the interval $t[-0.014,-0.2]$: (a) Energies, (b) Fidelity, (c) Entropy and (d) Maximum local and normal components. \textcolor{red}{ In panel (d)}  full figures corresponds to maximum local components, while the corresponding empty figure refers to the normal component.}
\end{figure}

The approaching of levels in the spectrum suggests a polyad breaking effect, but it does not provide a precise information about the region where it takes place. In Fig. 3b the fidelity is displayed for the three states.   As a reference the avoided crossings are marked with vertical lines in gray. While states $27$ and $28$ show a sensitive behavior under this property, state $26$ presents a small change, very close to the harmonic limit  until the the first avoided crossing appears.  This is explained by the existing  competition between local and normal character of the eigenstates displayed in panel (d). The fidelity detects slope component changes which may appear near or at the crossings of the local-normal maximum  components, manifested along the transition. At the  locations of avoided crossings, the fidelity curves of the states $26$ and $27$ are interchanged indicating  state crossings.  A similar situation appears at the second point, where the states $27$ and $28$ are interchanged. 
These crossings appear along the transition  because we are in a high energy region, but at low energies where no crossings appear,  the properties displayed also detect the LNT in the same parametric region.

In Fig. 3c the entropy is exhibited for the three states. In the pure local limit the entanglement  and consequently the entropy is expected to vanish, this explains its small values near H$_2$O.  As we move to the CO$_2$ parameters there is an entropy change associated with the LNT with features closely related to the local maximum component. After the transition  the entropy of the states  tend to the harmonic limit, which corresponds to constant values of the entropy. Here the avoided crossings are also manifested.

 In Fig. 3d we present the square of the maximum component in both  local and normal bases. The local components correspond to filled figures. The state $27$ starts with an almost purely local character (0.95).  As $|t|$ increases the local character rapidly diminishes with a proportional increasing of entropy. A similar situation appears in the state $28$, although in this case a maximum and minimum appears, in accordance to the local maximum component behavior.
 In contrast,  the state $26$ does not present such change in the first part, but after the crossing a change of dominance appears and  detected by the fidelity. Hence fidelity and entropy reflects in different form the subtle changes in the character of the eigenstates.

A picture of the studied states can be obtained  by plotting the probability density distribution in the coordinate representation. In Fig. 4 the probability density is shown for the three states as a function of the parameter $t$.  The  crossing points are indicated with full circles.
Except for the state $26$, the other two states present an evident  local character with the parameters of water molecule. The state $26$ contains a mixed character, a feature reflected by the components in Fig. 3d. As $|t|$ increases,  the transformation to NM character becomes manifest.  This visual point of view however is quite imprecise since after the first crossing the  change in the probability densities stop being noticeable, in contrast to the fidelity which continue detecting changes. The analysis of the plots before and after the red circles (avoided crossings) shows clearly that a crossing of states takes place \cite{hari,gbo}. At the crossing points a small change in the probability densities is revealed, although in general it will depend on the energy region  as well as on the value of the parameter.

\begin{figure}
\begin{center}
\includegraphics[width = 1.0\columnwidth]{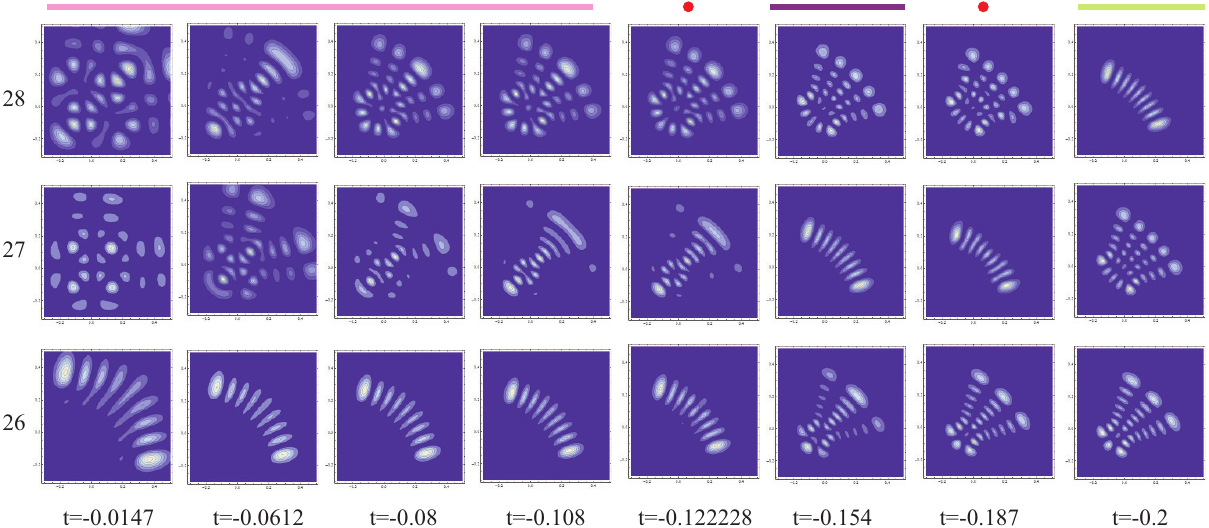}
\end{center}
\caption{Probability densities for the set of states $\{ 26,27,28 \}$ as a function of the parameter $x_g$. The avoided energy crossings are indicated with a red circles.}
\end{figure}

Finally in Fig. 5, the Poincar\'e sections for the  state $27$ are shown for different values of the parameter. Each plot is associated with the corresponding energy in Fig. 2, in such a way that it changes as we move to the normal limit represented by CO$_2$. In these plots we notice  that the LNT is manifested by a chaotic behavior. This is reasonable because the appearance of chaos has been associated with  lacking of preserved quantities like the polyad number \cite{kellmancaos,jung}. In the local limit we have integrable trajectories as well as in the normal limit. In the former case the polyad $P_L$ is preserved, while in the latter $P_N$ is a good quantum number. Hence it is in between, where the transition takes place, manifested with the appearance of  chaos. In Fig. 5 only the state $27$ is analyzed because the other two states are so close in energy that classically they do not provide additional information.   Although the chaotic transient regime associated with Poincar\'e sections is  energy dependent, we can establish that a global chaotic regime indicates the LNT in more precise terms, something that we cannot say just with the condition (\ref{condi12}). Notice that while the analysis  leading to the condition (\ref{condi12})  was based on the harmonic limit,  the study in terms of Morse oscillators display the transition regime as suggested.

\begin{figure}[t]
\begin{center}
\includegraphics[width = 1.0\columnwidth]{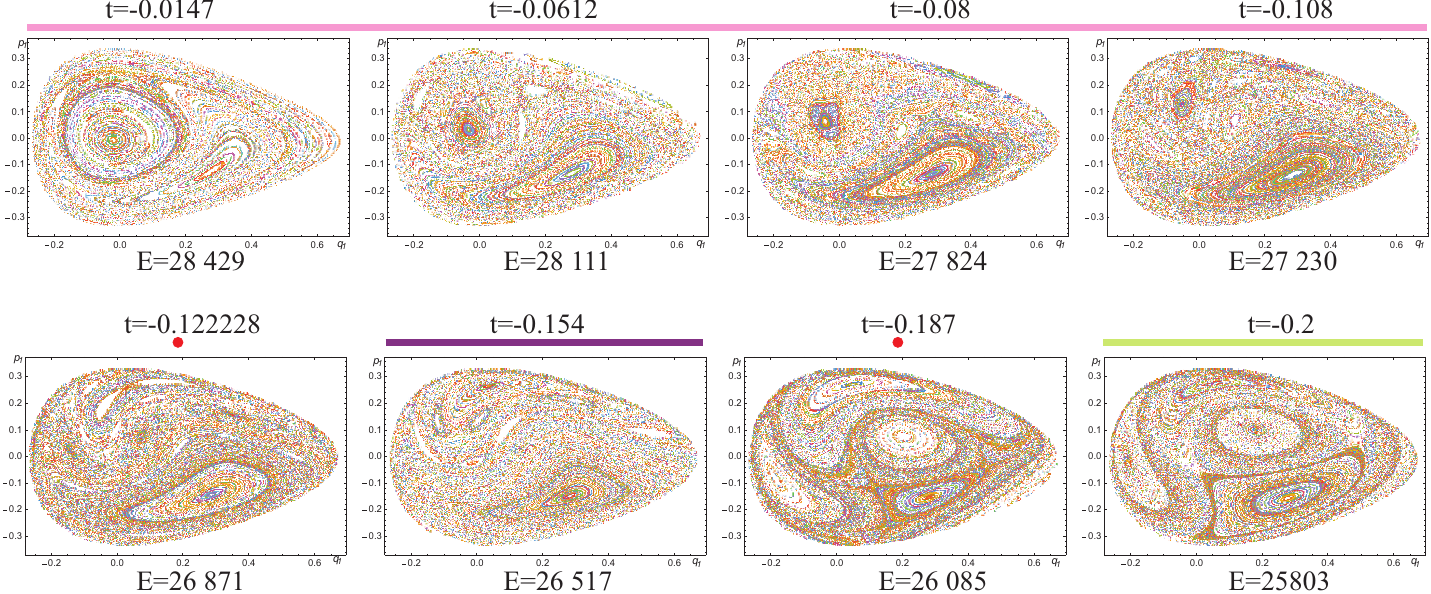}
\end{center}
\caption{Poincar\'e sections for the state $27$ as a function of the parameter $x_g$.}
\end{figure}

Appearance of chaos in a system of interacting Morse oscillators has been detected previously by considering   different  coupling strengths. Since working with exact interacting Morse oscillators involves the appearance of all the resonances \cite{brumer},   Poincar\'e sections are displayed in Ref.\cite{jung} to elucidate the appropriate approximation which preserve the essential features of the system. \textcolor{red}{ Coupling strength in the kinetic energy is related to masses and geometry of the system. Increasing this coupling may be interpreted either by a   mass ratio  and/or geometry modifications. In this sense the appearance of chaos in 
Refs.~\cite{terasaka,jung,kellmancaos} may be interpreted as parametric transformations  in the same direction as the presented analysis. Our work however address the polyad breaking phenomenon in search of regions where a system cannot be considered neither normal nor local, which is precisely associated with the appearance of chaos.}

\section{ Conclusions}

\textcolor{red}{In this work we have identified the LNT taking into account that the condition $P_L=P_N$ is not valid in the whole range of  force and structure constants}. The LNT has been studied using a 1D parametric form of the Hamiltonian for two interacting Morse oscillators. This system was chosen because it carries the main ingredients to successfully describe molecules with a LM behavior.   The  parametric form was based on the transformation from water to carbon dioxide molecule, through the set $\{x_g, x_f, \kappa \}$ of parameters in linear form. Although in our analysis the   transition can be identified with  a specific range of the parameter $x_g$, the transition features  vary with the energy.  This analysis differs from the previous studies of LMT in the sense that we are evaluating the range of parameters where the local force constants cannot be estimated using a local model, a region where the polyad $P_L$ stop being preserved.

The transition was studied using  several properties which  proved to furnish significant physical insight into the molecular behavior.  The fidelity is a  sensitive property that detects slope changes in the maximum components of the eigenstates. On the other hand the entropy reflects the local character of the eigenstates by its increasing when the local character diminishes. These properties are complementary and provide the  relevant interval of the transition,  but also the detailed transformation undergone by the  eigenstates.

The probability densities were also analyzed during the transition,
 \textcolor{red}{which it is not manifested in  all the states}.
 From the correlation diagram we detect avoiding   crossings. These crossings  do not establish a signature of the transition and appear as a consequence of the high energy region analyzed. States of lower energy undergo the transition in the same region without presenting any crossing. 
Finally a semiclassical feature of our analysis is the identification of a clear LNT interval through the appearance of chaos coinciding with the the abrupt changes in fidelity and entropy.
\acknowledgments
This work was supported by DGAPA-UNAM under project IN109113 and CONACyT with reference number 238494. First author is also grateful for the scholarship (Posgrado en Ciencias Qu\'imicas) provided by CONACyT, M\'exico.

\end{document}